\documentclass[journal,comsoc]{IEEEtran}

\usepackage[T1]{fontenc}
\usepackage[final]{graphicx}
\usepackage{amsmath}
\usepackage{amssymb}
\usepackage{subfig}
\usepackage{bm}
\usepackage{bbm}
\usepackage{subfig}
\usepackage{algorithm}
\usepackage{algpseudocode}
\usepackage{cite}
\usepackage{caption}

\newtheorem{thm}{Theorem}

\newtheorem{lem}{Lemma}

\title{Dynamic Spectrum Access using Stochastic Multi-User  Bandits}
\author{Meghana~Bande,~\IEEEmembership{Member,~IEEE,}
        Akshayaa~Magesh,~\IEEEmembership{Student~Member,~IEEE,}
        and~Venugopal~V.~Veeravalli,~\IEEEmembership{Fellow,~IEEE}
\thanks{M. Bande was with the Department
of Electrical and Computer Engineering, University of Illinois at Urbana-Champaign,
IL, 61820 USA. She is now with Qualcomm Technologies Inc., Bridgewater, NJ 08807 USA (email: mbande@qti.qualcomm.com)}
\thanks{A. Magesh and V.V. Veeravalli are with the Department of Electrical and Computer Engineering, University of Illinois at Urbana-Champaign, IL, 61820 USA. (email: amagesh2@illinois.edu, vvv@illinois.edu)} 
\thanks{This research was supported by the US NSF SpecEES under grant number 1730882, through the University of Illinois at Urbana-Champaign.}}

\begin{document}
\maketitle
\begin{abstract}
A stochastic multi-user multi-armed bandit framework is used to develop algorithms for uncoordinated spectrum access. In contrast to prior work, it is assumed that rewards can be non-zero even under collisions, thus allowing for the number of users to be greater than the number of channels. The proposed algorithm consists of an estimation phase and an allocation phase. It is shown that if every user adopts the algorithm, the system wide regret is order-optimal of order $O(\log T)$ over a time-horizon of duration $T$. The regret guarantees hold for both the cases where the number of users is greater than or less than the number of channels. The algorithm is extended to the dynamic case where the number of users in the system evolves over time, and is shown to lead to sub-linear regret.
\end{abstract}

\section{Introduction}
Dynamic spectrum access has emerged to address the problem of spectrum under-utilization caused by treating the frequency spectrum as a fixed commodity. 
We study the spectrum sharing paradigm in which all the users are treated equally i.e., there is no distinction between primary or secondary users.
We model the system as a stochastic multi-user multi-armed bandit (MAB) problem \cite{bubeck12survey} where the channels correspond to the arms of the bandit similar to the model considered in \cite{bande19icnc, Anand11, AvnerM14mega, RosenskiSS15mc, avner2015learning, kaufmann, magesh2019arxiv,  fu2009learning, liu2008cooperation,magesh2019}. The interference in the system is captured through the reward observed by each user. We propose a decentralized algorithm that  leads to efficient channel access and achieves sub-linear regret with time when employed by each user independently.

Stochastic multi-armed bandits have been used to model dynamic spectrum access extensively in literature. Multi-armed bandits with coordination between users was studied in \cite{fu2009learning}, \cite{liu2008negotiation}. We are more interested in the uncoordinated  stochastic multi-armed bandit model investigated in \cite{Anand11}, \cite{AvnerM14mega} and \cite {RosenskiSS15mc}. The algorithm in \cite{Anand11} achieves optimal regret but restricts the number of users to be lesser than the number of channels. The algorithms in \cite{AvnerM14mega} and \cite {RosenskiSS15mc} provide only high probability bounds on the expected regret.  

All of these approaches assume that when more than one user tries to transmit on the same channel simultaneously (commonly referred to as a collision), the colliding users receive zero reward, due to which the number of users in the system is constrained to be less than the number of channels. Hence, these approaches are not applicable to the case where the number of users is greater than the number of channels. In our model, 
%
each user receives a reward depending on which channel they choose and the number of other users that choose the channel at the same time. We consider a more general setting where the users can receive a non-zero reward when more than one user accesses the same channel with the reward for each user decreasing as a function of the total number of users on the channel. The work in \cite{tekin2011performance} also considers a setting with non-zero rewards on collisions, and provides guarantees for the expected time to converge to an optimal allocation when there is no explicit communication among the players. However, they assume that users have knowledge of the total number of users occupying their channel at any given time.

On any given channel, we assume that the reward obtained is a random variable that is drawn from a distribution that depends on the number of users on the channel. For example, the instantaneous reward could be the rate achieved by the user on the channel which may decrease due to interference from other users accessing the channel. The decrease in the reward observed by the user as a function of number of users depends on the system parameters, e.g., the distance between the users and transmission protocol (e.g., hybrid ARQ). 
In our model, the users do not communicate with each other. However, we do make the mild assumption that a low-bandwidth broadcast channel is available to the users for time synchronization (see also \cite{RosenskiSS15mc,nieminen2009time,avner2015learning}).  

A preliminary version of this work was considered in \cite{bande19icnc}, in which an algorithm was presented with guarantees of constant regret with high probability. Upon a more careful examination of the assumptions on the reward distributions of the arms and the clustering algorithm (Algorithm 2 in \cite{bande19icnc}), we believe that the results of the Theorem 2 in \cite{bande19icnc} do not hold under the assumptions stated. In our current work, we have been able to remove these assumptions and avoid the clustering approach altogether, and we present an algorithm achieving the order optimal regret of $O(\log T)$. We show that when each user employs our algorithm, the accumulated system-wide regret is $O(\log T)$ where $T$ is the time-horizon. We then consider the more realistic scenario where the number of users in the system changes over time with minor restrictions on the rate at which users can enter and leave the system. For this dynamic setting, we show that our algorithm can be easily extended to achieve sub-linear regret.

\section{System Model and notation}\label{sec:sysmod}

Let $K$ be the number of users in the system. We initially assume that the users have unlimited data for transmission. In a more realistic setting, users may become active or inactive depending on their transmission needs; our dynamic setting (see Section \ref{sec:dynm}) covers this scenario. Each user can choose one among $M$
channels for transmission. We assume that each user has prior knowledge of $M$. The assumption of known $M$ is reasonable if the spectrum partition is enforced and fixed.

We model the system as a stochastic multi-user multi-armed bandit (MAB) system with $K$ users and $M$ arms (channels). 
In each time slot $t$, let ${\cal A}_{t,j}$ denote the set of channels available to user
$j$. User $j$ chooses a channel $a_{t,j}\in {\cal A}_{t,j}$  based on the reward history according to a certain policy and receives a reward. The reward on each arm depends on the number of users who have chosen
the arm. Let $\mathbf{k}_t = [k_t(1), \ldots,k_t(M)]$, with $k_t(m)$ denoting the number of users on channel  $m$ at time $t$, and $\sum_{m=1}^{M}k_t(m) = K$. Let the reward received by user $j$  at time $t$  is a function of the channel chosen $a_{t,j}$ and the number of users on the channel $k_t(a_{t,j})$, and is denoted by $r(a_{t,j}, k_t(a_{t,j}))$. Note that the reward $r(a_{t,j}, k_t(a_{t,j}))$ depends on the channel chosen by all users, and this dependence is captured through $k(a_{t,j})$. The reward $r$ is normalized to lie in the interval $[0,1]$.

We model the system as a stochastic multi-user MAB system with $K$ users and $M$ arms (channels).  Each user can choose one among $M$
channels for transmission, where we allow for the possibility that $K\geq M$. 
As mentioned earlier, we assume that the reward observed decreases with the number of users transmitting
on the same channel. 
Let $\mu(m,k(m))$ denote the mean reward on channel $m$ when the number of users on the channel is $k(m)$, \textit{i.e.}, $\mu(m,k(m)) = \mathbb{E}[{r(m,k(m))}]$. We assume that $\mu(m,k(m))$ is 0 for $k(m)=N+1$, where $N$ is a constant that depends on the system. This imposes a restriction on the number of users in the system since $K$ cannot be greater than $MN$.

We define the expected regret in the system as, $\mathbb{E}[{R(T)}] = $
\begin{align*}
T\sum_{m=1}^{M}k^{\ast}(m)\mu(m,k^{\ast}(m)) - \sum_{t,j}\mathbb{E}[r(a_{t,j},k_t(a_{t,j}))]
\end{align*}
where 
$$ \mathbf{k}^{\ast}= \mathop{\arg\max}_{\mathbf{k}} \sum_{m=1}^{M}k(m)\mu(m,k(m)) $$ 
over all feasible $\mathbf{k}$ such that $\mathbf{k} \in \{\mathbb{N}\cup 0\}^{1\times M}$ and $\sum_{m=1}^{M} k(m)=K$. Note that $\mathbf{k}^{\ast}$ corresponds to the optimal number of users on each channel.

Since the reward distributions of the arms do not vary across the users, the optimal configuration (users occupy channels according to $\mathbf{k^*}$) does not depend on the channel allocated to any particular user. The mean reward of one channel may be greater than the others, and in order to ensure that one user does not monopolize a channel for an extended period of time, we impose the following condition. For each user, transmission on a particular channel takes place for a maximum of $T_x$ time slots, after which the user releases the channel for at least $T_x$ time slots before attempting to access the same channel. This notion of fairness does not interfere with the optimality of the system.

Let $J_1 = \sum_{m=1}^{M}k^{\ast}(m)\mu(m,k^{\ast}(m))$ be the system reward for the optimal configuration, and $J_2$ the system reward for the configuration that achieves the next possible lower value for system reward. In our algorithm we assume that we have access to a lower bound on the value $$\Delta = \frac{J_1 - J_2}{2 MN}.$$ Note that $\Delta > 0$, even though there might be multiple optimal configurations $\mathbf{k^*}$ achieving the system reward $J_1$. Such an assumption is usually required for the analysis of multi-user MABs when communication between users is not allowed. In the case where a bound on $\Delta$ is not known, the method of increasing exploration phases and eliminating sub-optimal matchings \cite{kaufmann} can be used to develop an algorithm that does not require the knowledge of $\Delta$.

In order for the users to get estimates of mean rewards of the arms, we assume that the users have unique IDs from $1$ to $K$ at the beginning of the algorithm. This assumption is required in order to devise a simple exploration phase when no assumptions on the reward distributions of the arms are made. Since the users have access to a low-bandwidth broadcast channel, unique IDs for users from $1$ to $K$ and the value of $K$ can be broadcast to the users at the beginning of the algorithm. In the dynamic case where users can enter and leave the system, the value of $K$ can be broadcast at the beginning of each epoch.

\section{Policy for Decentralized Multi-User Multi Armed Bandits}\label{sec:alg}

The decentralized policy for each user (Algorithm \ref{sto:main}) proceeds in epochs, with the number of epochs being $L$ over a horizon of length $T$. Each epoch consists of two phases. The first is an estimation phase during which each user estimates the mean reward as a function of the number of users $k(m)$ on each channel $m$. Using these estimates, each user then computes an optimal configuration of number of users on the channels. The second is an allocation phase where the users align themselves according to the optimal system configuration. We show that our algorithm leads to sub-linear regret of $O(\log T)$ where $T$ is the time-horizon.

The estimation phase is for user $j$ to obtain estimates of mean rewards (denoted by $\hat{\mu}_j(m,n)$) of arms $m \in [M]$ for all $n \in [N]$. This phase proceeds for a fixed number of time units in every epoch. Since every user has an unique ID, and the total number of users $K$ is known, the users simply sample each arm $m$, for each value of $n$ from $1$ to $N$, for $T_0 = \frac{1}{2\Delta^2}$ time units.

Note that if there is more than one optimal configuration in the system, the algorithm can dictate how the users make a decision about the estimate. For example, in the event of multiple configurations with same reward, the users choose the one with increasing number of users on the channels.

We use Algorithm \ref{sto:alloc} to construct an efficient allocation for which the regret does not grow with time when all the users have the correct estimate for the optimal configuration. During each epoch $\ell$, this phase proceeds for $2^\ell$ time units. At the beginning of this phase in each epoch, users occupy channels in order of their IDs corresponding to their estimated  $\mathbf{\hat{k}^*}$. In order to ensure fairness, the users switch channels in a round robin fashion after every $T_x$ time units. This parameter can be selected according to the size of each epoch. Note that, if $|\hat{\mu}^j(m,n)-\mu(m,n)|\leq \Delta$ for all $j \in [K], m \in [M]$ and $n \in [N]$, we have that $\mathbf{\hat{k}^*} = \mathbf{k}^*$, and the system does not incur regret during the allocation phase.

\begin{algorithm}[h]
  \caption{}
  \label{sto:main}
  \begin{algorithmic}
 \For {epoch $\ell = 1$ to $L$}
 \State {\bfseries Estimation Phase}: Run Algorithm \ref{estimate} 
  
  \State Calculate $\mathbf{\hat{k}}^{\ast}$ from $\hat{\mu}(m,n),n\in [N],K$
\State {\bfseries Allocation Phase}: Run Algorithm \ref{sto:alloc} 
    \EndFor
  \end{algorithmic}
\end{algorithm}

\begin{algorithm}
   \caption{Estimation Phase}
   \label{estimate}
\begin{algorithmic}
   \For{$n = 1$ to $N$}
   \State Users divide in groups of $n$ in order of IDs and complete groups of $n$ play arms $1$ to $M$ for $T_0$ time units
   \State If final group is incomplete, it is completed with users from group 1 and completed group plays arms 1 to $M$ for $T_0$ time units
   \EndFor
\end{algorithmic}
\end{algorithm}{}

\begin{algorithm}[h]
  \caption{Allocation phase}
    \label{sto:alloc}
  \begin{algorithmic}
  \For{$t = 1$ to $2^\ell$}
  \State Users occupy channels in order of IDs according to estimated optimal allocation $\mathbf{\hat{k}^*}$
  \State After every $T_x$ time units, users switch channels in round robin fashion
  \EndFor
  \end{algorithmic}
  \label{alg:permute}  
\end{algorithm}

\section{Analysis}\label{sec:anal}

We show that after the estimation phase, with high probability, each user has the correct estimates for $\mu(m,k(m))$. More precisely, each user $j$ computes estimates $\hat{\mu}^j(m,n)$ such that $|\hat{\mu}^j(m,n)-\mu(m,n)|\leq \Delta$ with high probability, for $n \in [N]$.

\begin{lem}\label{sto:mu}
During each epoch $\ell$, for each user $j$, channel $m$ and number of users $n \in [N]$ on the channel, when the estimation phase is carried out with $T_0 = \frac{1}{2\Delta^2}$, with probability at least $1-e^{-\ell}$,
\[|\hat{\mu}^j(m,n)-\mu(m,n)|\leq \Delta.\]
\end{lem}

The proof follows from Hoeffding's inequality \cite{hoeff}.

We now present the upper bound on the expected regret incurred by the users employing Algorithm \ref{sto:main}.

\begin{thm}
The expected regret incurred by employing Algorithm \ref{sto:main} is given by
\begin{equation}\label{regret}
    \mathbb{E}[R(T)] \leq \frac{K^2MN}{2\Delta^2} \log T + \frac{2K^2MN}{e-2}.
\end{equation}
\end{thm}
\begin{IEEEproof}
Let $L$ denote the number of complete epochs in time horizon $T$. By construction of the algorithm, we have that $L < \log T$. Let the regret incurred during the estimation phase of all epochs be denoted by $R_e$ and the regret incurred during the allocation phase be denoted by $R_a$. 

The estimation phase in each epoch proceeds for at most $\frac{KMN}{2\Delta^2}$ time units. Thus, 
\begin{equation}
    R_e \leq \sum_{i=1}^L \frac{K^2 M N}{2\Delta^2} \leq \frac{K^2 M N}{2\Delta^2} \log T  .
\end{equation}

Note that regret is incurred in the allocation phase only in the case when there exists some user $j \in [K]$, some channel $m \in [M]$ and some $n \in [N]$ such that $|\hat{\mu}_j(m,n) - \mu_j(m,n)| > \Delta$. We have from Lemma \ref{sto:mu}, that the probability of this event is upper bounded by $KMN e^{-\ell}$. Thus we have that 
\begin{equation}
    R_a \leq \sum_{i=1}^L K^2MN \frac{2}{e}^\ell \leq \frac{2K^2MN}{e-2}.
\end{equation}

Therefore, the expected regret for Algorithm \ref{sto:main} for a time horizon $T$ is given by
\begin{equation}
    \mathbb{E}[R(T)] \leq \frac{K^2 M N}{2\Delta^2} \log T + \frac{2K^2MN}{e-2} \sim O(\log T).
\end{equation}
\end{IEEEproof}
\section{Dynamic case}\label{sec:dynm}

We now extend the results to a dynamic system with a changing number of users. The key idea is to run Algorithm  \ref{sto:main} repeatedly in super-epochs, each consisting of epochs described in Section \ref{sec:alg}. In order to obtain a sub-linear regret bound, we restrict the number of users entering and leaving the system until time $t$, denoted by $\kappa_t$, to be sub-linear. Let $\kappa_t$ be $O(t^{\zeta})$, where $\zeta<\frac{1}{2}$. We note that this is different from \cite{RosenskiSS15mc} where the time-horizon is fixed and known, and there is also a restriction on when users can enter or leave the system. 

Each user considers some known time $\tau$ which is greater than the estimation phase ($\frac{(MN)^3}{2\Delta^2}\log \tau$ time-units) and runs Algorithm  \ref{sto:main}. After time $\tau$, the user continues to use Algorithm  \ref{sto:main} with a super-epoch of length 2$\tau$, then $3 \tau$, and so on. Let $K_t$ denote the number of active users at time $t$, where $K_t\leq MN$. The resulting algorithm is given in Algorithm \ref{sto:dyn}.  
\begin{algorithm}[htb]
  \caption{Dynamic Allocation}
  \label{sto:dyn}
  \begin{algorithmic}
 \For{$\tau \sum_{i=1}^{r}i\leq T \leq \tau \sum_{i=1}^{r+1}i$}
 \State Run Algorithm  \ref{sto:main} 
 \EndFor
  \end{algorithmic}
\end{algorithm}

We now show that if all the users employ Algorithm \ref{sto:dyn}, the system-wide regret is sub-linear in $T$ when $\zeta<\frac{1}{2}$. We emphasize that the users do not need to know the time-horizon $T$ to achieve sub-linear regret.

\begin{figure}
    \centering
    \includegraphics[scale= 0.9]{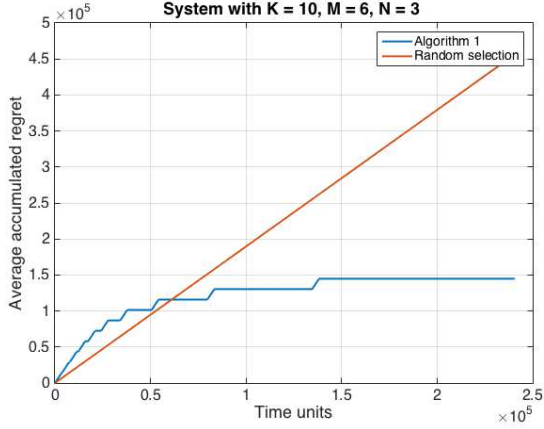}
    \caption{Average accumulated regret over $100$ runs. }
    \label{fig:regret}
\end{figure}
\begin{thm}
The expected regret $\mathbb{E}[R(T)]$ after running Algorithm \ref{sto:dyn} until time $T$  where $\tau \sum_{i=1}^{r}i\leq T \leq \tau \sum_{i=1}^{r+1}i$ is upper bounded by
\[(MN)^3\left[\sqrt{2T}C + \frac{(\sqrt{2T}+1)\log 2T}{4\Delta^2}\right] + MN \kappa_T \sqrt{2T},\]
where $C = \frac{\log \tau}{2\Delta^2} + 1.4$ and, $\mathbb{E}[R(T)]\sim O(T^\frac{1}{2}\log T+\kappa_T T^\frac{1}{2})$.
\end{thm}
\begin{IEEEproof}
Note that $r\leq \sqrt{2T}$, and recall that $K_t\leq MN$.

Let ${\cal E}_r$ denote the set of super-epochs until $r$ during which at least one user enters or leaves the system. Note that $|{\cal E}_r|\leq \kappa_T$. 
Let $R_i$ denote the regret accumulated in super-epoch $i$. In super-epochs where no users enter or leave the system, the regret is bounded according to Theorem \ref{sto:main}, and in super-epochs in ${\cal E}_r$, the regret accumulates through the entire super-epoch. 
The expected regret in super-epochs with change is given by: 
\begin{eqnarray*}
\sum_{i\in {\cal E}_r}\mathbb{E}[R_i] &\leq& MN\sum_{i\in {\cal E}_r}\text{Length of super-epoch $i$ }\\
&\leq&  MN|{\cal E}_r| r\tau
\leq MN\kappa_T r\tau. 
\end{eqnarray*}

The regret in super-epochs with no change is bounded using Theorem \ref{sto:main} as
\begin{align*}
\sum_{i\in [r]\backslash {\cal E}_r}\mathbb{E}[R_i]& 
\leq \sum_{\ell=1}^{r}\mathbb{E}[R_{\ell}]  \leq (MN)^3\left(\sum_{\ell=1}^{r}\frac{\log r\tau}{2\Delta^2}  + \frac{r}{e-2}\right) \\
   &\leq  (MN)^3\left[r\left(\frac{\log \tau}{2\Delta^2}  + \frac{1}{e-2}\right) + \frac{r+1\log r}{2\Delta^2}\right]  \\
   &\leq (MN)^3\left[\sqrt{2T}C + \frac{(\sqrt{2T}+1)\log 2T}{4\Delta^2}\right], 
 \end{align*}   
where $C = \frac{\log \tau}{2\Delta^2} + 1.4$, which is a constant.

The regret up to time $T$ bounded as follows:
\begin{align*}
\mathbb{E}[R(T)] & = \sum_{i\in [r]\backslash {\cal E}_r}\mathbb{E}[R_i]+\sum_{i\in {\cal E}_r}\mathbb{E}[R_i]\\
   \leq (MN)^3 &\left[\sqrt{2T}C + \frac{(\sqrt{2T}+1)\log 2T}{4\Delta^2}\right] + MN \kappa_T\sqrt{2T}\tau.
 \end{align*}   
Thus,  $\mathbb{E}[R(T)]\sim O(T^\frac{1}{2}\log T+\kappa_T T^\frac{1}{2})$, and if  $\kappa_T$ is $O(T^{\zeta})$ with $\zeta<\frac{1}{2}$, we have sub-linear regret.
\end{IEEEproof}
\section{Experiments}

We consider a system with $K=10$ users and $M=6$ channels with $N=3$. The reward distributions are chosen to be uniform with a variance of $0.01$, and means between $0$ and $1$. The performances of Algorithm \ref{sto:main} and an algorithm where the users choose channels uniformly at random are compared in Fig. \ref{fig:regret}. It can be seen from the figure that the regret incurred by the naive random selection algorithm is linear, whereas the regret incurred by Algorithm \ref{sto:main} is sub-linear. Algorithm \ref{sto:main} performs worse initially due to a shorter allocation phase in each epoch compared to estimation phase. Note that the allocation phase of epoch $\ell$ proceeds for $2^\ell$ time units, and we can see from the flat regions of the plot that the regret incurred during the allocation phase is zero with high probability.  
\section{Conclusion}
We developed algorithms for uncoordinated spectrum access within the framework of stochastic multi-armed bandits. We allowed for the users to receive non-zero rewards on collisions, and for the number of users to be greater than the number of channels. In this setup, we presented an algorithm that achieves order-optimal system regret of $O(\log T)$. We also presented an algorithm that achieves sub-linear regret for the dynamic case where the number of users evolves over time. It is of interest to extend the results in this paper to the case of heterogeneous reward distributions across arms; some initial results in this direction are explored in \cite{magesh2019, magesh2019arxiv}.

\bibliographystyle{IEEEtran}
\bibliography{ref}

\end{document}